# Probabilistic annotation of protein sequences based on functional classifications

Emmanuel D Levy[1,3], Christos A Ouzounis*[1], Walter R Gilks[2] and Benjamin Audit*[1,4]

Address: [1]Computational Genomics Group, The European Bioinformatics Institute, EMBL Cambridge Outstation, Cambridge CB10 1SD, UK, [2]Medical Research Council Biostatistics Unit, Institute of Public Health, Cambridge CB2 2SR, UK, [3]Computational Genomics Group, MRC Laboratory of Molecular Biology, Hills Rd, Cambridge CB2 2QH, UK and [4]Laboratoire Joliot-Curie and Laboratoire de Physique, CNRS UMR5672, Ecole Normale Supérieure, 46 Allée d'Italie, 69364 Lyon Cedex 07, France

Email: Emmanuel D Levy - elevy@mrc-lmb.cam.ac.uk; Christos A Ouzounis* - ouzounis@ebi.ac.uk; Walter R Gilks - wally.gilks@mrc-bsu.cam.ac.uk; Benjamin Audit* - Benjamin.Audit@ens-lyon.fr

* Corresponding authors

## Abstract

**Background:** One of the most evident achievements of bioinformatics is the development of methods that transfer biological knowledge from characterised proteins to uncharacterised sequences. This mode of protein function assignment is mostly based on the detection of sequence similarity and the premise that functional properties are conserved during evolution. Most automatic approaches developed to date rely on the identification of clusters of homologous proteins and the mapping of new proteins onto these clusters, which are expected to share functional characteristics.

**Results:** Here, we inverse the logic of this process, by considering the mapping of sequences directly to a functional classification instead of mapping functions to a sequence clustering. In this mode, the starting point is a database of labelled proteins according to a functional classification scheme, and the subsequent use of sequence similarity allows defining the membership of new proteins to these functional classes. In this framework, we define the Correspondence Indicators as measures of relationship between sequence and function and further formulate two Bayesian approaches to estimate the probability for a sequence of unknown function to belong to a functional class. This approach allows the parametrisation of different sequence search strategies and provides a direct measure of annotation error rates. We validate this approach with a database of enzymes labelled by their corresponding four-digit EC numbers and analyse specific cases.

**Conclusion:** The performance of this method is significantly higher than the simple strategy consisting in transferring the annotation from the highest scoring BLAST match and is expected to find applications in automated functional annotation pipelines.

## Background

The gap between the growth rate of biological sequence databases and the capability to characterise experimentally the roles and functions associated with these new sequences is constantly increasing [1]. This results in an accumulation of raw data that can lead to an increase in our biological knowledge only if computational characterisation tools are developed. We focus here on the annotation of protein function. A generic approach to this problem consists of transferring the annotation from sequences of known function to uncharacterised proteins [2]. The transfer mechanism might be subdivided in two steps: (i) to establish the list of known proteins with significant sequence similarity to the uncharacterised sequence; (ii) to select the known sequence(s) from which the annotation is transferred [3]. The first step is usually performed with sequence alignment tools such as FASTA [4] or BLAST [5]. When sensitivity is critical, alternative tools such as PSI-BLAST [6] and hidden Markov models [7] can be used. Finding homologous proteins can also be accomplished using alignment-independent sequence comparison tools, which have been developed to overcome the limitation arising from the assumption of contiguity between homologous segments [8,9]. Then, the challenge is the selection of true homologues from the list of similar sequences. Most of the above tools provide a score measuring the degree of similarity between the sequences compared. A simple criterion to single out a homologue is to choose the most similar sequence i.e. the highest scoring sequence. More elaborate methods have been designed to enhance the precision and reliability of the annotation process. These rely on the combination of the annotations of more than one homologue [10-13] or, for example, on semantic analyses of annotation lines [14].

This type of annotation process relies on the assumption of a strong relationship between protein sequence and function. This hypothesis is generally fair [15] even though many studies have demonstrated the existence of counter-examples that can lead to annotation errors [16-19]. Two major origins of errors can be distinguished: (i) the short listed homologous protein(s) have a different function from the sequence to be annotated (failure of the sequence-function paradigm or error in the homology search); (ii) the transferred annotations were themselves not correct (transfer of database errors). The second type of errors along with the iterative usage of annotation transfer gives rise to the specific problem of error propagation when newly annotated sequences are included in the reference database used for the homology search. Recent studies have shown that dramatic consequences on the reliability of database annotations are likely to arise from this process [20]. In order to improve our control on these two types of errors, it would be very useful to associate a measure of reliability to the annotations obtained. In this way, we might limit the introduction of new errors and limit their propagation by not admitting the transfer of the less reliable annotations.

In this work, we address this issue by developing a probabilistic framework to the homology-based annotation process. Our approach relies on the usage of a reference dataset where protein sequences are classified into functional classes. Here, an annotation is a membership to a functional class, thus, function sharing is evident. The possibility for a protein to perform a particular function is then assessed based on its similarity relationships with all protein sequences known to perform this function; it enables for instance to take into consideration both the presence and the absence of similarity. This possibility is used during the training step of machine-learning approaches for sequence annotation, which relies on the availability of a classified reference dataset [21-23]. Note that most other methods proposed to date map function to proteins by first "clustering" proteins based on sequence similarities and second combining the functional description of the characterised proteins to propose a description for the uncharacterised sequences. The present approach inverts this process by mapping sequences to a functional classification instead of mapping functions to a sequence clustering. Following this idea, we propose a method to build *correspondence indicators* (CIs) between sequences and functional classes. Then, we explore two Bayesian annotation frameworks based on the comparison of the CIs of a sequence of unknown function with the observed CIs for the reference protein sequences. This framework provides probabilities for a sequence to belong to the different functional classes. We advocate the use of these probabilities as a direct measure of the reliability of annotations.

To validate both probabilistic methods for automatic annotation, we applied them to the well-established classification of enzymes. Our results show that both methods allow distinguishing proteins whose annotation is reliable from the others. At the highest level of reliability, the two methods predict the four EC digits with a very low error rate (~0.002) for 90.6% and 96.0% of enzymes respectively. We compared these results with the simple strategy consisting in transferring the EC number of the BLAST best hit. Our best method has an error rate half that of the best-hit strategy at the same coverage level.

## Results
### Defining correspondence indicators
Given a functional classification, annotating a new protein consists in establishing to which functional class or classes it belongs. To approach the problem we defined a *Correspondence Indicator* (CI) between the new protein and each of the functional classes, and second, formulated a classification scheme based on these indicators. This section is devoted to the first point, whereas the second one will be treated in the following section.

Using the bit-scores of sequence alignments (See Methods), we can imagine many different scoring strategies to measure this correspondence. For instance, we could use the number of hits (with a bit-score above a given threshold), or the best bit-score between the new protein and the functional class members. Alternatively, we might choose to compromise between the two above options by taking the sum of the bit-scores between the new protein and the class members.

Here, we propose a measure that unifies these three strategies. Let $\Omega_1, ..., \Omega_n$ symbolise the set of $n$ functional classes with respective sizes $N_1, ..., N_n$. We denote $S_{c,d}$ the BLAST bit-score between two proteins $c$ and $d$. Then, we define the CI $Y^{(\alpha)}_{\Omega_j}(c)$ parameterised by $\alpha \in [0, +\infty)$, between a new protein $c$ and the class $\Omega_j$ as follows:

$$Y^{(\alpha)}_{\Omega_j}(c) = \sum_{d \in \Omega_j} S^{\alpha}_{c,d}, \qquad (1)$$

where the sum is taken over bit-scores $S_{c,d}$ greater than a given threshold $S_0$, for $c \neq d$. $Y^{(\alpha)}_{\Omega_j}(c)$ measures the strength of the homology relationship between the new protein $c$ and the family $\Omega_j$ as the sum of the pairwise bit-scores to the power $\alpha$ between $c$ and all class members. The possibility to choose the parameter $\alpha$ allows modulating the relative weight of each hit with a class: the relative weight of hits with larger bit-scores increases with $\alpha$. Using $\alpha = 0$, all the hits have the same weight and $Y^{(\alpha=0)}_{\Omega_j}(c)$ is simply the "number of hits" of $c$ with $\Omega_j$. When $\alpha = 1$, the weight of each hit is its bit-score and $Y^{(\alpha=1)}_{\Omega_j}(c)$ is the "sum of the bit-scores". Finally, as $\alpha \to \infty$ only the hit with the largest bit-score counts and $Y^{(\alpha \to \infty)}_{\Omega_j}(c)$ reduces to $\max_{d \in j}^{\alpha}(S_{c,d})$, which is equivalent to the "best bit-score" scoring strategy. (For large value of $\alpha$, it can be convenient to simply define $Y^{(\alpha \to \infty)}_{\Omega_j}(c)$ as the best bit-score with class $\Omega_j$: $Y^{(\alpha \to \infty)}_{\Omega_j}(c) = \max_{d \in j}(S_{c,d})$). Thus, the choice of $\alpha$ enables a continuous variation between the strategies where only the number of hits or only the bit-score of the best hit counts. To our knowledge, such a parametric approach to sequence search metrics has not been proposed previously. Note that a more sophisticated combination of the CIs obtained for different $\alpha$ values could take advantage of various aspect of information captured by each of these $\alpha$ values.

### Different strategies of annotation
*Best correspondence indicator strategy*

Given a fixed value for $\alpha$, the simplest classification scheme is to assign the new protein $c$ to the class $\Omega_{j_0}$ that maximises the CI. For $\alpha = 0$, it is the functional class with the greatest number of hits with $c$. When $\alpha \to \infty$, this amounts to the class containing the sequence closest to $c$: a simple "best hit" strategy of annotation. Note that different values of $\alpha$ may result in a different classification of $c$.

*Estimating the probability for a protein sequence to belong to a functional class: an univariate Bayesian approach*

A limitation to the "best CI" strategy of annotation is the lack of a reliability assessment for the functional assignments. To overcome this limitation, we propose to estimate, independently for each of the functional classes, the probability $P(c \in \Omega_j \mid Y^{(\alpha)}_{\Omega_j}(c))$ for a protein $c$ drawn at random, to belong to class $\Omega_j$ given $Y^{(\alpha)}_{\Omega_j}(c)$ i.e. we estimate probabilities knowing one variable (indicator) only. Using Bayes theorem, we can show [See Additional file 1, Section S2] that this conditional probability can be estimated by:

$$\hat{P}(c \in \Omega_j \mid Y^{(\alpha)}_{\Omega_j}(c)) = \frac{N_{\Omega_j}(Y^{(\alpha)}_{\Omega_j}(c) \pm \lambda)}{N(Y^{(\alpha)}_{\Omega_j}(c) \pm \lambda)}. \qquad (2)$$

where $N_{\Omega_j}(Y^{(\alpha)}_{\Omega_j}(c) \pm \lambda)$ and $N(Y^{(\alpha)}_{\Omega_j}(c) \pm \lambda)$ are respectively, the number of proteins truly belonging to class $\Omega_j$ and the number of proteins from the entire dataset, whose correspondence indicator with class $\Omega_j$ is comprised in $[(Y^{(\alpha)}_{\Omega_j}(c) - \lambda), (Y^{(\alpha)}_{\Omega_j}(c) + \lambda)]$. This mechanism is illustrated in Additional file 1, Figures S1(a). $\lambda$ is fixed for the annotation of each new protein such that the total number of sampled proteins $N(Y^{(\alpha)}_{\Omega_j}(d) \pm \lambda)$ is always equal to 10. This can be viewed as an adaptive smoothing of the data: $\lambda$ is increased until the interval $[(Y^{(\alpha)}_{\Omega_j}(c) - \lambda)$,

**Table 1: Performance of the Univariate Bayesian annotation approach.** Re-annotation of the filtered ENZYME database with the univariate Bayesian approach. Since we systematically sample 10 enzymes to calculate the probabilities for a protein to belong to each functional class (See Different strategies of annotation), probabilities can only take one of the following eleven values: 0, 0.1, ..., 0.9, 1. We report for each assignment probability level and globally the number of correct annotations, the number of annotation errors and the corresponding error rate and coverage of the database.

|  | Univariate Bayesian approach | | | | | | | | | | | |
|---|---|---|---|---|---|---|---|---|---|---|---|---|
| **Assignment probability** | 0 | 0.1 | 0.2 | 0.3 | 0.4 | 0.5 | 0.6 | 0.7 | 0.8 | 0.9 | 1 | TOT |
| **Correct annotations** | 84 | 109 | 103 | 99 | 119 | 177 | 252 | 302 | 437 | 726 | **25387** | 27795 |
| **Annotation errors** | 27 | 15 | 5 | 11 | 13 | 23 | 41 | 29 | 31 | 45 | **53** | 293 |
| **Error rate (%)** | 24.3 | 12.1 | 4.6 | 10.0 | 9.8 | 11.5 | 14.0 | 8.8 | 6.6 | 5.8 | **0.21** | 1.04 |
| **Coverage (%)** | 0.4 | 0.4 | 0.4 | 0.4 | 0.5 | 0.7 | 1.0 | 1.2 | 1.7 | 2.7 | **90.6** | 100.0 |

($Y_{\Omega_j}^{(\alpha)}(c) + \lambda$)] contains a predetermined quantity of information (10 proteins).

*Determining the most likely functional class of a protein sequence: a multivariate Bayesian method of annotation*

In the previous approach, we assessed the membership of a new protein to a functional class using only the CI with this class. Because this process is performed independently for each class, it allows several probabilities to be close to 1. In such circumstances functional assignment is ambiguous. To improve the control on these cases, we propose to estimate the probability $P(c \in \Omega_j | \{Y_{\Omega_1}^{(\alpha)}(c) ... Y_{\Omega_n}^{(\alpha)}(c)\})$ of a new protein $c$ to belong to $\Omega_j$ knowing the set $\{Y_{\Omega_1}^{(\alpha)}(c) ... Y_{\Omega_n}^{(\alpha)}(c)\}$ of its CIs with all the functional classes i.e. we estimate probabilities based on multiple variables (indicators). Using Bayes theorem, we can show [See Additional file 1, Section S3] that:

$$\hat{P}(c \in \Omega_j | \{Y_{\Omega_1}^{(\alpha)}(c)...Y_{\Omega_n}^{(\alpha)}(c)\}) = \frac{N_{\Omega_j}(B(\{Y_{\Omega_1}^{(\alpha)}(c)...Y_{\Omega_n}^{(\alpha)}(c)\},r))}{N(B(\{Y_{\Omega_1}^{(\alpha)}(c)...Y_{\Omega_n}^{(\alpha)}(c)\},r))}. \quad (3)$$

Estimating this probability amounts to consider the $n$-dimensional space of CIs and to look in that space what is the functional composition of the proteins that have their position within the sphere $B(\{Y_{\Omega_1}^{(\alpha)}(c) ... Y_{\Omega_n}^{(\alpha)}(c)\}, r)$ of radius $r$, centred at $\{Y_{\Omega_1}^{(\alpha)}(c) ... Y_{\Omega_n}^{(\alpha)}(c)\}$. In other words, we count the number of proteins within the sphere $B(\{Y_{\Omega_1}^{(\alpha)}(c) ... Y_{\Omega_n}^{(\alpha)}(c)\}, r)$ that truly belong to class $\Omega_j$. The ratio between this number and the total number of proteins in the sphere is $\hat{P}(c \in \Omega_j | \{Y_{\Omega_1}^{(\alpha)}(c) ... Y_{\Omega_n}^{(\alpha)}(c)\})$. This mechanism is illustrated in Additional file 1, Figure S1(b).

As previously for $\lambda$, $r$ is determined for each protein such that the total number of proteins sampled $N(B(\{Y_{\Omega_1}^{(\alpha)}(c) ... Y_{\Omega_n}^{(\alpha)}(c)\}, r))$ is always 10. Note that this method amounts to find the 10 closest proteins from the reference dataset to the point $\{Y_{\Omega_1}^{(\alpha)}(c) ... Y_{\Omega_n}^{(\alpha)}(c)\}$ in the CI space. The logic behind this adaptive methodology is that the local density of proteins in the CI space can be highly variable depending on the average level of homology between proteins in each functional class. Hence, using constant value for $\lambda$ and $r$ is not adequate. In this framework, it does not make sense to attempt classifying proteins into classes with less than 10 members. The particular choice of 10 corresponds to a trade off between precision (the higher the number of proteins in the neighbourhood, the higher the precision of probability calculations; see the caption from Table 1) and coverage i.e. the number of EC classes considered (see Methods).

*Determining the optimal correspondence indicator*

The freedom of choice of the parameter $\alpha$ in the CI $Y_{\Omega_j}^{(\alpha)}(c)$ (Eq.(1)) allows us to combine in different ways the bit-scores of the alignments of protein $c$ with the proteins of the class $\Omega_j$. The choice of $\alpha$ enables a continuous variation between the strategies where only the number of hits above the threshold $S_0$ ($\alpha = 0$) or only the bit-score of the best hit ($\alpha \to \infty$) counts (See Defining correspondence indicators). To optimise the parameters $\alpha$ and $S_0$, we re-annotated each enzyme (See Methods) using the best CI strategy (See Different strategies of annotation) with different combinations of their values. In Figure 1, we report the number of annotation errors $E(\alpha, S_0)$ for each combination of these parameters. At a fixed value of $S_0$, we observe that the higher $\alpha$, the lower the number of errors.

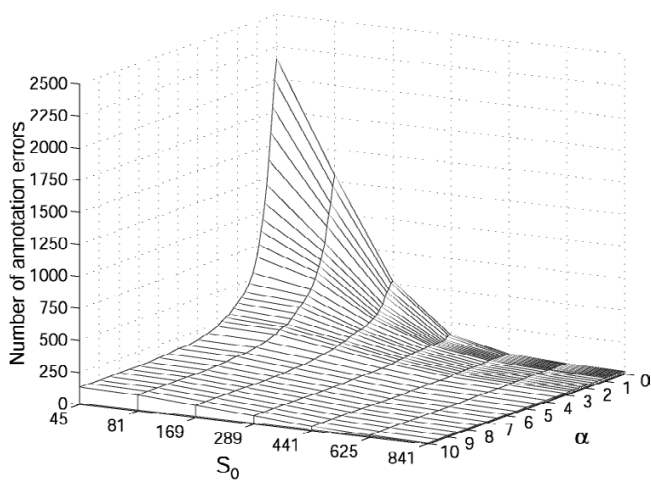

**Figure 1**
**Number of re-annotation errors**. Number of annotation errors $E(\alpha, S_0)$ made during the re-annotation of the 28088 enzymes of the filtered ENZYME database (See Methods) using the best CI strategy (See Different strategies of annotation) as a function of the parameter $\alpha$ and cut-off $S_0$ (Eq. (1)).

Moreover, the difference between the number of errors at $\alpha = 0$ and at $\alpha = 10$ soars for small $S_0$ values (20 folds at $S_0 = 45$: $E(0,45) = 2156$ and $E(10,45) = 122$). This effect is due to the poor specificity of alignments having a small bit-score. With no cut-off ($S_0 = 0$), all sequences hit one another and thus, for $\alpha = 0$ all the functional assignments are made to the largest class. This illustrates that the choice of the $\alpha$ value can be critical, and that for small $\alpha$ values, the sensitivity to $S_0$ is higher. At a fixed value of $\alpha$, by increasing the cut-off $S_0$ we minimise the number of errors as expected, but simultaneously the coverage of annotated proteins drops significantly (Fig. 2). Indeed, a protein cannot be annotated if all its hits have a bit-score below the cut-off. This shows that if we do not want to lose too much coverage, there is very little latitude on $S_0$.

By minimising the number of errors to determine the optimal value for $\alpha$, we conclude that the best bit-score strategy ($\alpha \to \infty$) is the one which best describes the relation between an enzyme and its functional class. Moreover, given the weak sensitivity to $S_0$ for $\alpha \to \infty$, we choose the smallest value $S_0 = 45$ for the threshold in order to maximise the coverage. Then, from now on, the only CI we will be using is $Y_{\Omega_j}^{(\alpha \to \infty)}(c)$ for $S_0 = 45$, denoted $Y_{\Omega_j}(c)$. The best CI strategy with $\alpha \to \infty$ reduces to a BLAST *best-hit strategy* (See Different strategies of annotation). Its performance depends only on the threshold $S_0$ that also directly controls the coverage (Fig. 2). It will serve as a reference to compare to the results obtained with the two probabilistic approaches. In this way, given the same input (BLAST pairwise bit-scores), we will assess the improvement in annotation quality obtained by an efficient usage of a functional classification on the reference dataset. Recently developed database search methods exploit sequence profiles and often outperform BLAST in terms of sensitivity for the detection of remote homologies. However, this increase in sensitivity usually comes at the expense of specificity, which is the most important feature in the present set up. Indeed, most enzymes have easily detectable homologies with sequences from their EC class.

### Re-annotation with the univariate Bayesian approach

The univariate Bayesian approach allows estimating the probabilities for an enzyme to belong to a particular EC class $\Omega_j$, given only $Y_{\Omega_j}(c)$ (the maximum bit-score with this class). To annotate an enzyme using this method, we derive probability estimates for its membership in each functional class (Eq.(2)) and assign it to the class for which the probability is largest. We re-annotated all enzymes of the reference dataset *via* this approach, using the leave-one-out method (See Methods). The results of this re-annotation are presented Table 1. There is a total annotation error rate $r = 0.010$ (293 errors), which is more than twice as large as for the re-annotation of the reference dataset by the best-hit strategy (Fig. 2: $r = 0.0045$, 126 errors). However, we can take advantage of the assignment probabilities that do not exist for the best-hit strategy and that are a direct measure of the confidence we have in an assignment. Considering the large proportion of proteins (90.6%) that are annotated with the highest confidence (assignment probability of 1), we notice that $r = 0.0021$ (53 errors), two fold smaller than for the best-hit strategy. The reduction of the error rate remains very significant even if we take into account the coverage of annotation: at the same coverage, the best-hit strategy leads to a rate of error of 0.0034 which is 1.5 larger (Fig. 2). To achieve the same rate of error with the best-hit strategy the coverage drops to 54% (Fig. 2). Interestingly, for assignment probabilities smaller than 1, the error rate dramatically increases ($r > 0.05$). These results demonstrate that flagging annotations with the assignment probabilities allows us to filter out likely errors. Finally, we note in Table 1 that 84 proteins are reported to be correctly annotated with an assignment probability equals to 0. These proteins only hit their true class and so, can be assigned only to it. However, because their CI with their class falls in a range of values with only protein from other classes, the assignment probability is equal to zero (Eq.(2)).

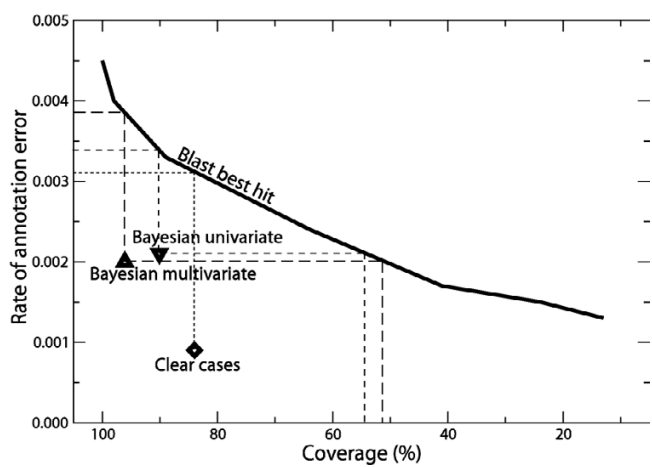

**Figure 2**
**Re-annotation error rate**. Rate of annotation error as a function of the coverage for the re-annotation of the 28088 enzymes of the filtered ENZYME database (See Methods). The full line corresponds to the *best-hit strategy* (See Determining the optimal correspondence indicator); the curve was obtained by performing the re-annotation for different values of the threshold $S_0$ between 45 (100% coverage by definition of the filtered ENZYME database) and 841. ($\nabla,\Delta$) correspond to the univariate and multivariate Bayesian methods at the highest confidence level (P = 1, Tables 1 and 2). ($\Diamond$) corresponds to the "clear cases" identified by the univariate Bayesian method (P = 1 and second highest probability equals to 0; see Re-annotation with the univariate Bayesian approach).

In this mode of automatic annotation, the probabilities of membership of a protein to each functional class are estimated independently, allowing for two or more probabilities to be significant e.g. 1 and 0.8. In principle, this property permits to assign a protein to more than one functional class. Nevertheless, if proteins can belong to one functional class only, as for the set of enzymes considered here (See Methods), these situations correspond to ambiguous cases that are more likely to lead to annotation errors than instances where proteins have only one significant probability. Indeed, out of the 25387 enzymes annotated with an assignment probability of 1 (Table 1), 23655 have their second highest probability equal to 0 (data not shown). For these "clear cases", the error rate is significantly reduced to r = 0.0009 (21 errors) which is 3 times smaller than the error rate for the maximum bit-score strategy at the same annotation coverage (Fig. 2; r = 0.0031 at 84% coverage). This result strongly suggests that taking into account simultaneously the CIs with all functional classes can lead to significant improvement in the annotation process. This approach is investigated in the next section.

*Re-annotation with the multivariate Bayesian method*
We now explore a multivariate Bayesian method taking into account all CIs concurrently. More precisely, each protein is mapped to a point in an *n*-dimensional space where each dimension corresponds to one of the *n* possible functional classes. In this space, the coordinates of a protein are the CIs $Y_{\Omega_j}(c)$ (maximum bit-score) with each family. The probabilities for a protein to belong to each functional class are estimated simultaneously according to the functional class of the 10 proteins of the reference dataset closest to the unclassified protein in this space (See Different strategies of annotation and Additional file 1, Fig. S1(b)). Note that compared to the univariate Bayesian approach, protein neighbourhood is determined globally, just once. As a result, the sum of all the probabilities is always 1; there cannot be more than one significant class membership probability (P~1) per protein.

We re-annotated all enzymes of the reference dataset *via* this method (See Methods, Table 2). Compared with the univariate approach, we note a decrease of the global error rate (r = 0.0079 vs. 0.010). At the highest annotation confidence (assignment probability of 1), we observe a significant increase of the annotation coverage (96.0% vs. 90.6%) concomitant with a stable error rate (r = 0.0020 and 53 errors vs. r = 0.0021 and 53 errors). The error rate at the highest confidence level is half that of the best-hit strategy for the same coverage. We observe that to achieve a similar error rate the coverage of the best-hit strategy would dramatically drop to 51% (Fig. 2). Interestingly, the assignment probabilities closely match the empirical error rates. For instance, for the set of enzymes annotated with an assignment probability of 0.7, we measure an error rate of 0.242 ($\approx$1-0.7).

*Comparing the two Bayesian annotation strategies*
The two Bayesian methodologies differ significantly on the coverage of the database of enzymes annotated at the maximum level of reliability (probability 1): 90.6% (25440/28088) for the univariate approach in contrast with 96.0% (26951/28088) for the multivariate method. This increase of coverage actually associated with a constant number of errors (53) corresponds to 1511 more correct annotations in favour of the multivariate method (Tables 1 and 2). This is due to the fact that the multivariate Bayesian method regards a protein sequence as a single point in the CI space while the univariate Bayesian approach considers the orthogonal projection on each CI axis separately. Figures 3(a) and 3(b) propose two examples to illustrate the consequences of this difference.

*Exploring the CI space for EC classes 2.3.1.61 (Dihydrolipoamide S-succinyltransferase) and 2.3.1.12 (Dihydrolipoamide S-acetyltransferase)*

Focusing on protein O31550 [Swiss-Prot:O31550] from EC 2.3.1.12, we note Figure 3(a) that its CIs (best bit-scores) with both EC classes are similar (231 on the Y-coordinate with EC2.3.1.12 and 225 on the X-coordinate with EC 2.3.1.61). To calculate the probabilities to belong to each EC classes with the multivariate Bayesian method, we look at the functional distribution of the proteins closest to O31550 in the CI space (See Different strategies of annotation, Eq.(3)). This process is represented by the dotted circle in Figure 3(a); it leads to $P_{2.3.1.12}$ = 0.7 and $P_{2.3.1.61}$ = 0.3 and, thus, to a correct annotation of O31550. By contrast, when annotating this protein with the univariate Bayesian approach, these probabilities are calculated independently (See Different strategies of annotation, Eq.(2)). $P_{2.3.1.12}$ falls to 0 because on the EC2.3.1.12 axis, around bit-score 231 (box to the right), we sample only proteins belonging to EC 2.3.1.61. In the same manner, for EC 2.3.1.61 around bit-score 225 (box on top), we observe only one protein out of 10 that truly belongs to EC 2.3.1.61 so that $P_{2.3.1.61}$ = 0.1. Hence, we wrongly assign O31550 to EC 2.3.1.61 but with a very low assignment probability $P$ = 0.1.

*Exploring the CI space for EC 1.6.5.3 (NADH dehydrogenase (ubiquinone)) and EC 1.6.99.5 (NADH dehydrogenase (quinone))*

There is also strong sequence similarity between proteins from these two EC classes and there exists a quite well defined "boundary" that is densely populated (Fig. 3(b)). Very clearly the projections on the CI axes intrinsic to the univariate approach tend to mix the 804 proteins from the two EC classes leading to poor performances (at P = 1, r = 0.014 for 44.2% coverage) whereas the multivariate method can adapt to the boundary and leads to improved performances (at P = 1, r = 0.0028 for 90.0% coverage). These cases clearly exemplify that the projections on the CI axes can have great influence on the probability calculation and may result in annotation errors. It also shows that the multivariate method outperforms the univariate approach because of its ability to adapt to the shape of the boundary between functional classes in the CI space.

**Analysing the origins of annotation errors**

The proposed Bayesian annotation strategies optimise the exploitation of the functional information carried by CIs built upon sequence similarity clues (BLAST bit-scores). We explore examples of the failure of these clues leading to annotation errors when using the multivariate Bayesian method.

*Annotation errors between Glyceraldehyde 3-phosphate dehydrogenases*

Proteins from classes EC 1.2.1.12 and EC 1.2.1.59 catalyse the same reaction (Glyceraldehyde 3-phosphate dehydrogenation) but EC 1.2.1.12 proteins are NAD-dependent while EC 1.2.1.59 proteins can use both NAD and NADP as cofactors. As illustrated in Figure 3(c), there exists strong cross-similarity between sequences from these two classes but each class tends to occupy a separate part of the CI space so that annotation can be done accurately. We note four exceptions: four proteins from EC 1.2.1.59 (black triangles; [Swiss-Prot:O09452, Swiss-Prot:O34425, Swiss-Prot:P80505, Swiss-Prot:Q48335]) are closer to the EC 1.2.1.12 cloud than to the other EC 1.2.1.59 proteins in the CI space and thus are wrongly re-annotated as EC 1.2.1.12 proteins. The erroneously re-annotated EC 1.2.1.59 sequence O34425 is *Bacillus subtilis* gapB protein. Protein gapA [Swiss-Prot:P09124], also from *B. subtilis* belongs to class EC 1.2.1.12. It was shown that gapA can acquire the gapB activity with only two amino acids mutations (D32A and L187N) [24]; actually, gapB possesses these mutations. Therefore, a reasonable hypothesis is that gapA and gapB originate from a gene duplication event followed by divergent evolution. From the topology of Figure 3(c), it is possible that similar scenarios apply to the three other "misplaced" EC 1.2.1.59 sequences. Here, functional specialisation can be achieved with only a few modifications at specific sites. General alignment tools like BLAST do not capture the higher significance of mutations at these sites compared to alterations at other sites; this leads to annotation errors difficult to avoid with automatic general-purpose tools.

*Annotation errors between two-sector ATPases*

Another interesting example of annotation errors comes from the classes EC 3.6.3.14 and EC 3.6.3.15, both of which contain transporting two-sector ATPases, the former transporting $H^+$ and the latter $Na^+$. In the CI space, the two clouds of points marking the proteins from these classes exactly overlap (data not shown) i.e. CIs based on BLAST bit-scores do not capture any sequence specificity distinguishing the two EC classes (the two classes are associated with the same 5 PROSITE patterns [25]). EC 3.6.3.15 being much less populated than EC 3.6.3.14 (16 members and 1252 members, respectively), this particular topology results in the 16 EC 3.6.3.15 sequences to be wrongly assigned to EC 3.6.3.14 with a high confidence (P = 1 in the 16 cases) because a large majority of their neighbours in the CI space belongs to EC 3.6.3.14. More generally, when CIs do not allow the distinction of two classes then we expect most sequences to be assigned to the larger class with an assignment probability equal to the relative size of this class. Hence, unless one class is greatly larger than the other one, assignment probabilities will be significantly smaller than 1 allowing us to filter out these specious annotations. In other words, the 16 erroneous annotations of EC 3.6.3.15 proteins originate from a class size effect. In most situations class sizes are of the same order and such a local topology of the CI space leads to easily detectable annotations errors (low assignment probabilities). This example of annotation errors actually explains how, by scanning the local configuration of the CI space, the Bayesian strategies can avoid a number of errors.

**Table 2: Performance of the Multivariate Bayesian annotation method.** Re-annotation of the filtered ENZYME database with the multivariate Bayesian method. Since we systematically sample 10 enzymes to calculate the probabilities for a protein to belong to each functional class (See Different strategies of annotation), probabilities can only take one of the following eleven values: 0, 0.1, ..., 0.9, 1. We report for each assignment probability level and globally the number of correct annotations, the number of annotation errors and the corresponding error rate and coverage of the database.

| | Multivariate Bayesian method | | | | | | | | | | | |
|---|---|---|---|---|---|---|---|---|---|---|---|---|
| **Assignment probability** | 0 | 0.1 | 0.2 | 0.3 | 0.4 | 0.5 | 0.6 | 0.7 | 0.8 | 0.9 | 1 | TOT |
| **Correct annotations** | 0 | 0 | 0 | 0 | 9 | 35 | 109 | 116 | 188 | 511 | 26898 | 27866 |
| **Annotation errors** | 0 | 0 | 0 | 5 | 10 | 37 | 34 | 37 | 29 | 17 | 53 | 222 |
| **Error rate (%)** | - | - | - | 100.0 | 52.6 | 51.4 | 23.4 | 24.2 | 13.4 | 3.2 | **0.20** | 0.79 |
| **Coverage (%)** | 0.0 | 0.0 | 0.0 | 0.0 | 0.1 | 0.3 | 0.5 | 0.5 | 0.8 | 1.9 | **96.0** | 100 |

*Example of annotation error propagation*
In the present work, we considered the annotations associated to the sequences in the ENZYME database to be exact. Nevertheless, analysis of the origins of annotation errors using visual representations of the CI space as shown Figure 3, revealed peculiar configurations of the sequence-function relationship. A close investigation of these cases allowed us to identify three clear annotation errors. Figure 3(d) provides an example of error identification. Protein P94598 [Swiss-Prot:P94598] is annotated as a member of EC 1.4.1.3 (NAD(P)-utilizing glutamate dehydrogenase) but the multivariate Bayesian method assigned it to EC 1.4.1.4 (NADP-specific glutamate dehydrogenase) with an assignment probability of 1. Indeed, P94598 is close to a group of EC 1.4.1.4 proteins in the CI space. Tracing the source of this annotation, we noted that its strong CI value with EC1.4.1.4 originated from a strong sequence similarity with protein P95544 [Swiss-Prot:P95544] annotated as EC 1.4.1.4. By checking the publication associated with the annotation of P95544, we noted that this protein was wrongly annotated and actually belongs to EC 1.4.1.3 [26]. Correcting this database annotation error, the CI value of P94598 with EC 1.4.1.3 increases while its CI value with EC 1.4.1.4 decreases so that in fact the multivariate Bayesian method correctly classifies it to EC 1.4.1.3 (Fig. 3(d)). Interestingly, this example provides an illustration of an annotation error susceptible to propagate [20]. The correction of the annotation of P95544 was submitted to the ENZYME database and is expected to be included in future releases. Another example comes from P17692 [Swiss-Prot:P17692] that we classify as EC 2.4.1.19 (cyclomaltodextrin glucanotrans-ferase) in disagreement with its database annotation: EC 3.2.1.1 (alpha-amylase). Actually, the EC 2.4.1.19 activity of P17692 has been described in the literature [27]. In addition, we found that Q11119 [Swiss-Prot:Q11119] (EC 3.1.2.14, oleoyl-[acyl-carrier protein] hydrolase) should be annotated as EC 3.1.2.15 (ubiquitin thiolesterase). The experts of ENZYME have validated these two annotation errors and corrected the corresponding database entries.

**Discussion**
The maintenance of various aspects of protein function is intricate due to the inhomogeneity of the sequence-function relationship. For example, 60% of EC classes with more than 2 members could not be perfectly discriminated by sequence similarity at any BLAST threshold [28]. Moreover, the 4 (or first 3) EC digits were systematically identical only above 80% (or 50%) sequence identity in structural alignments, while at the other end of the spectrum, the preservation of the 4 EC digits was observed at as low as 16% identity [29]. Consequently, the threshold below which sequence similarity should not be considered for annotation transfer at a given confidence level should in general be determined for each functional class independently. However, it is typically set in a uniform manner. In sharp contrast, the two Bayesian methods developed here take into account how functional classes are distributed locally in the relevant part of the CI space or along CI axes and assign a low probability where the sequence-function relationship is ambiguous.

Interestingly, with both Bayesian approaches, a large majority of proteins have been re-annotated with an assignment probability of 1 (Tables 1 and 2). In the case of the multivariate Bayesian method, it means that for 96.0% of the enzymes of our dataset their 10 nearest enzymes in the CI space have the same EC number. Also, at the fourth level of the EC hierarchy, 255 classes out of 589 (43%) are isolated i.e. the 10553 proteins out of 28088 (38%) belonging to these classes have no BLAST hit (above threshold $S_0 = 45$) with the proteins of the other classes. This illustrates that there exists a high level

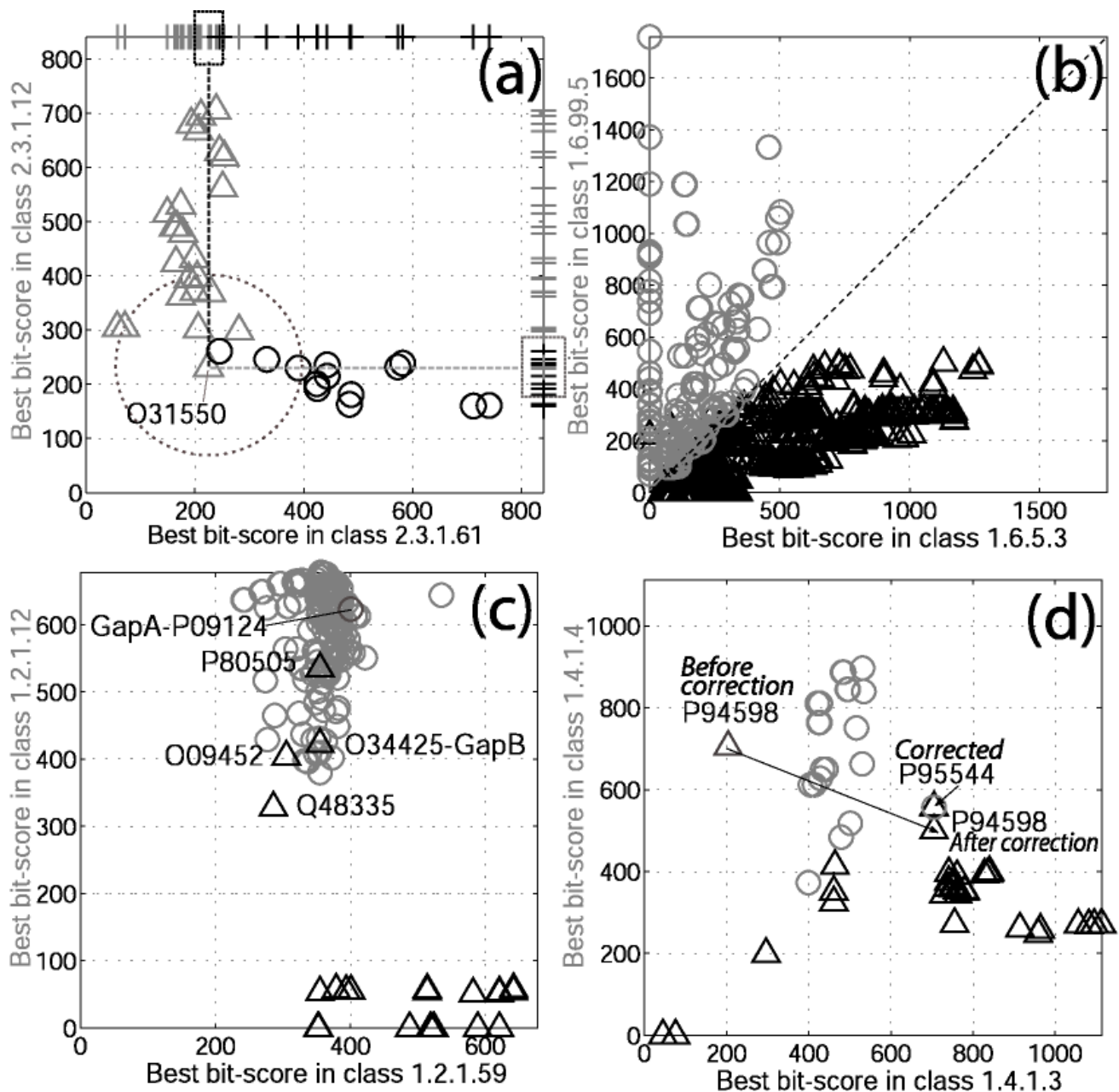

#### Figure 3
**Examples of topology in the CI space.** In 4 cases where there is a strong cross similarity between sequences belonging to two different EC classes, we plot for each protein of these classes a point whose 2 coordinates are the CIs of its sequence with the two functional classes (BLAST best-hit with the corresponding EC class). (a): EC 2.3.1.61 (black circles) and EC 2.3.1.12 (grey triangles); crosses on top and at right correspond to the projection on the CI axes; the dotted circle (boxes on top and to the right) marks the limit of the sampling regions used to annotate O31550 with the multivariate (univariate) Bayesian method [See Additional file 1, Fig. S1]. (b): EC 1.6.5.3 (black triangles) and EC 1.6.99.5 (grey circles). (c): EC 1.2.1.59 (black triangles) and EC 1.2.1.12 (grey circles). (d): EC 1.4.1.3 (black triangles) and EC 1.4.1.4 (grey circles); the arrow shows the change of position of protein P94598 in the CI space when the annotation of P95544 is corrected from EC 1.4.1.4 to EC 1.4.1.3 (See Analysing the origins of annotation errors).

of clustering of enzymes sharing their four EC digits in the CI and sequence spaces. Thus, for the filtered ENZYME database we considered in this analysis (enzymes catalysing one reaction only and EC categories with more than 11 members; see Methods), CI based on sequence similarity is a meaningful clue to predict the full EC code. In contrast, considering EC digit conservation based on pairwise sequence comparison, it was found that a good practical rule was to transfer 2 EC digits above 15% sequence identity [29]. There is no contradiction here. Essentially, when considering sufficiently populated EC classes, for most sequences we find very close homologues within their class allowing a clear functional annotation. This property of large EC classes also explains why the optimal CIs are obtained for $\alpha \to \infty$ (See Determining the optimal correspondence indicator) i.e. why the optimal CIs reduce to the best BLAST bit-score with each class while the number of hits is not taken into consideration (See Defining correspondence indicators): the important property in the sequence-EC class relationship is that the EC class contains at least one highly similar sequence to the query sequence under study. This situation also clarifies the reason for the good performance of the simple BLAST best-hit strategy for the tested data set (error rate smaller than 0.0045; Fig. 2). A priori, the well-specified clustering of sequences belonging to the same class cannot be generalised to other classifications of proteins, so depending on the sequence classification scheme under consideration it is important to measure the optimal $\alpha$ value. In situations where this value is small (i.e. when the number of hits is more significant than their scores), it is predictable that the difference between the performances of the Bayesian approaches and the simple BLAST best-hit method will be greatly increased.

## Conclusion

The importance of standardising the systems by which biological functions are described is now generally recognised [30]. This has opened up the possibility for high-throughput automatic retrieval of sequences based on functional characteristics. In the present work, we demonstrate the great potential offered by a classification of protein functions to improve the quality of sequence annotations. Indeed, the availability of such a functional classification allows the definition of measures of correspondence between a sequence and all functional classes i.e. it permits taking advantage of the complete set of similarity relationships of a query sequence with the sequences from a reference database. The automated Bayesian methodologies provide reliable information about the sequences whose assignment probability is large enough (in this work, P = 1) leaving behind the more "difficult" cases. In an annotation pipeline, these methodologies could be an efficient filter to focus the work of human experts on the more error prone cases [31]. Along the same lines, inconsistencies between automated annotation and database annotation could be used to highlight possible annotation errors [32]; in this context, visual representations like those presented in Figure 3 can be a useful tool for human experts.

An important aspect of this work is the construction of correspondence indicators between sequences and functional classes (Eq.(1)). Here, we used BLAST bit-scores for this process but the score from any pairwise protein comparison can be used instead e.g. structural comparison [33,34] or alignment-independent measures that can be computed from the primary sequence like length, word frequency, molecular weight or total charge [8,9,22]. Note that in principle, any measure of relationship between sequence and function can be used instead of CIs. In a previous study, it was shown that the simple BLAST best-hit approach outperformed three machine-learning methods based on alignment-independent features for the classification of enzymes within the EC hierarchy [22]. In contrast, the two Bayesian classifiers based on CIs outperform sequence similarity alone in term of sensitivity and specificity. This suggests that CIs could reveal themselves to be powerful features as input to machine-learning approaches for protein classification [21,23]. It remains to be seen whether the performance of CIs based on pairwise BLAST bit-scores is constant across various classification problems e.g. when there is only remote homology between class members [35].

The analytical development leading to CIs can be extended to construct a measure of correspondence between two functional classes that describes the degree of their overlap in the CI space (Fig. 3). Since a strong overlap indicates that two functional classes cannot be distinguished by the CIs, we can build an "adapted" functional classification by merging functional classes based on this new criterion. Interestingly, this amounts to empirically solve the problem of the extent of the functional annotation that can be transferred [29]. For example, EC 3.6.3.14 and EC 3.6.3.15 exactly overlap in the CI space (See Analysing the origins of annotation errors), this means that BLAST-based CIs simply do not differentiate these two types of transporting two-sector ATPases. It is more effective in an automated system to group these two classes in a Meta EC class "Na$^+$ or H$^+$ transporting two-sector ATPases" that we can reliably assign to. A key feature of the proposed methodologies is the quantification of the reliability of annotations; the assignment probability represents an attractive candidate, both versatile and compact, to qualify non-experimentally based annotations. In principle, it could be taken into account by the Bayesian annotation framework allowing its iterative usage without risking the propagation of annotation errors [20]. It is our hope that the Bayesian annotation strategies presented herein will contribute to more robust automatic annotation pipelines.

## Methods
### A database of enzymes
In the present work, we put forward a method of classification of uncharacterised proteins, based on their pattern of homology with a reference set of classified proteins. We validate this approach on a database of enzymes annotated by their four-digit EC number. Annotations and sequences have been retrieved using release 30 of ENZYME [36]) and release 41 of SWISSPROT [37]. We quantified the homology relationship between two proteins by the bit-score of the alignment between their sequence using BLAST with default parameters settings [6]. Query sequences were masked for low-complexity regions using CAST [38]. Where BLAST reports more than one significant hit between two sequences, we retain only the best bit-score. We performed a BLAST "all against all" comparison between enzymes and stored all pairwise best bit-scores greater than 45 (E-value cut-off of $10^{-5}$ for the database under consideration).

The tree-like structure of the EC nomenclature [See Additional file 1, Section S1] suggests that the EC classification defines a functional partition of enzymes. However, 1078 enzymes are classified into multiple EC classes. This can originate from overlaps in the definition of EC classes, or from multi-functional enzymes. In the present work, we do not take explicitly into account the possibility of multi-functional proteins. Hence, all enzymes with more than one EC number were discarded in order to obtain a reference dataset where the functional classification defines a partition of the protein sequence set. In addition, protein sequences annotated as "fragment" in SWISSPROT have not been considered. Ultimately, the probabilistic framework of annotation we developed requires a minimum number of proteins in each class for the functional assignments to be meaningful. We fixed this minimum number to 10 proteins and so ignored all classes containing less than 11 members (we re-annotate each enzyme using a leave-one-out method; see Validation by re-annotation). Finally, we removed the 215 sequences that did not present any hit in our database of local alignments. This defined the reference set of 28088 protein sequences used in the present analysis as well as their functional classification.

### Validation by re-annotation
In order to quantify the performance of the different annotation strategies presented above, they were applied to re-annotate the filtered ENZYME database using a leave-one-out procedure. This method consists in removing in turn each enzyme from the reference dataset and to re-annotate it as if it was a new enzyme of unknown activity. The so-obtained classification of enzymes was then compared to the original classification. For the two Bayesian methods, new enzymes were assigned to the functional class for which the estimated probability is the highest.

## Authors' contributions
All authors participated in the design of the study and writing of the manuscript. EDL implemented the methodology and performed the analysis. All authors read and approved the final manuscript.

## Additional material

**Additional File 1**
*contains a more detailed description of the EC nomenclature to complement section "A database of enzymes" and the full calculation leading to equations(2) and(3) along with a figure illustrating their meaning.*
Click here for file
[http://www.biomedcentral.com/content/supplementary/1471-2105-6-302-S1.pdf]

## Acknowledgements
We thank members of the Computational Genomics Group for comments, Kristian Axelsen for helpful exchanges, G. Akoun and S. Maslau for reading the manuscript. C.A.O. acknowledges additional support from IBM Research.